\documentclass[aps,prl,a4paper,amsmath,twocolumn,superscriptaddress]{revtex4-1}
\usepackage{graphicx}
\usepackage{bm}
\usepackage{color} 
\usepackage{soul}

\def\grad{\ensuremath{^{\circ}}}

\def\lina{PbCuSO$_{4}$(OH)$_{2}$}

\definecolor{petrol}{rgb}{0,0.5,0.5}
\definecolor{bluered}{rgb}{0.5,0,1}
\definecolor{darkblue}{rgb}{0,0,0.5}
\definecolor{darkgreen}{rgb}{0,0.5,0}
\definecolor{darkgrey}{rgb}{0.3,0.3,0.3}
\definecolor{grey}{rgb}{0.5,0.5,0.5}
\definecolor{malacol}{rgb}{0.12,0.69,0.44}
\definecolor{orangered}{rgb}{1,0.2,0.2}
\definecolor{darkyellow}{rgb}{1,.9,0.09}

\begin{document}
\title{Linarite -- a quasi-critical $J_1$-$J_2$ chain}

\author{Eron Cemal}
   \affiliation{Institut Laue-Langevin, CS 20156, 38042 Grenoble Cedex 9, France}

\author{Mechthild Enderle}
   \affiliation{Institut Laue-Langevin, CS 20156, 38042 Grenoble Cedex 9, France}

\author{Reinhard K. Kremer}
   \affiliation{Max-Planck Institute for Solid State Research, Heisenbergstrasse 1, 70569 Stuttgart, Germany}

\author{Bj\"orn F{\aa}k}
   \affiliation{Institut Laue-Langevin, CS 20156, 38042 Grenoble Cedex 9, France}

\author{Eric Ressouche}
   \affiliation{Universit\'{e} Grenoble Alpes, CEA, INAC, MEM, F-38000 Grenoble, France}

\author{Mariya V. Gvozdikova}
   \affiliation{Institut Laue-Langevin, CS 20156, 38042 Grenoble Cedex 9, France}

\author{Mike E. Zhitomirsky}
   \affiliation{CEA, INAC-PHELIQS, F-38000 Grenoble, France}

\author{Tim Ziman}
   \affiliation{Institut Laue-Langevin, CS 20156, 38042 Grenoble Cedex 9, France}
   \affiliation{LPMMC, UMR-5493, Universit\'e Grenoble Alpes and CNRS, 38042 Grenoble, France}

\date{\today}

\begin{abstract}
The mineral linarite, PbCuSO$_4$(OH)$_2$, is a spin 1/2 chain with frustrating nearest neighbor ferromagnetic and next-nearest neighbor antiferromagnetic exchange interactions. Our inelastic neutron scattering experiments performed above the saturation field establish that the ratio between these exchanges is such that linarite is extremely close to the quantum critical point between spin-multipolar phases and the ferromagnetic state. However, the measured complex magnetic phase diagram depends strongly on the magnetic field direction. The field-dependent phase sequence is explained by our classical simulations of a nearly critical model with tiny orthorhombic exchange anisotropy. The simulations also capture qualitatively the measured variations of the wave vector as well as the staggered and the uniform magnetizations in an applied field. 
\end{abstract}

\maketitle

Interacting spin-1/2 systems provide a rich source of remarkable discoveries due to the intricate nature of quantum many-body 
physics. While interacting semiclassical spins $S\gg 1/2$ tend to arrange themselves into periodic magnetic structures with a finite 
dipole moment at each site, the quantum character of $S=1/2$ spins may lead to disordered spin liquid ground states \cite{Zhou17,Savary17}.
Such quantum spin liquids are prone to appear in the vicinity of critical points between vastly different ground states \cite{Starykh15};  examples are  spin liquids at the boundary between dimerization and long-range magnetic order \cite{Sandvik07} and the multitude of chiral quantum spin liquids emerging between different types of long-range order on the kagome lattices \cite{Bieri15}.
Of particular interest are spin models with frustrated ferro- and antiferromagnetic couplings,
which have unusual spin-multipolar quantum phases stabilized by an external magnetic field in the proximity of a ferromagnetic ground state 
\cite{Sha06,Vek07,Kek07,Hik08,Sud09,Hei09,Shi09,Zhi10,Ued09,Bal16,Jan16}.
The quantum spin-multipolar phases have entangled spin multipoles extending over two or more lattice sites in addition to the incomplete uniform alignment of the spin dipole moments parallel to the field.

In spin-1/2 chains with nearest-neighbor (NN) ferromagnetic (FM) and next-nearest-neighbor (NNN) antiferromagnetic (AFM) interactions, the transverse spin-multipolar quasi-long range order is accompanied by  longitudinal spin-dipolar quasi-long-range order with an incommensurate propagation vector $k_{\rm{ic}}=\frac{1}{2p}(1-m)$, where $m$ is the uniform moment per site and $p=2,3, \ldots$ for quadrupolar, octupolar, \ldots spin tensors, respectively
\cite{Vek07,Kek07,Hik08,Sud09,Sat09}.
The dipolar correlations of the $p$-spin density wave (p-SDW) can be probed in neutron scattering experiments
as demonstrated in LiCuVO$_4$, where pair-correlations with incommensurate propagation vector $k_{\rm{ic}}=\frac{1}{4}(1-m)$ ($p=2$) and spins parallel to the field have been observed \cite{Mou12}. The existence of spin-multipolar bond order with $p>2$, however, is far less obvious and might depend on the presence or absence of the spin rotational $U(1)$ symmetry about the field direction \cite{Sud09,Bal16,Nis15}.
For instance, replacement of the $U(1)$ symmetry by two-fold $C_2$ rotations mixes
all odd magnon sectors preventing the distinction between the ordering of third-rank octupolar spin tensors and usual dipole ordering. In this context, the mineral linarite plays a particular role: It has been established that the ratio of NN-FM and NNN-AFM exchange \cite{Wil12,Rul17} is closer to the critical value than in LiCuVO$_4$ \cite{End05,End10}. In addition, the propagation vector does not follow the simple relation $k_{\rm{ic}}=\frac{1}{2p}(1-m)$ \cite{Wil16}, and the reported complex phase diagram \cite{Wil12,Wol12,Scha13,Pov16} has remained unexplained. 

In this Letter, we combine magnetization and neutron diffraction measurements with zero-temperature simulations of a $S=1/2$ model to show that a tiny orthorhombic anisotropy and weak interchain interactions explain the topology of the phase diagram of linarite for three orthogonal field directions as well as the variations of the ordering wave vector in high magnetic fields. 
Using inelastic neutron scattering (INS) in a strong magnetic field we determine the ratio of the NN and NNN exchange constants and establish unambiguously that linarite is indeed extremely close to the critical point.
These results imply that weak anisotropies have a dramatic influence on the field-dependent phase diagram and stabilize ordered dipolar states, raising doubts about the possibility of spin-multipolar ordering beyond quadrupoles in linarite, and in real materials in general.

\begin{figure}
\includegraphics[width=\columnwidth]{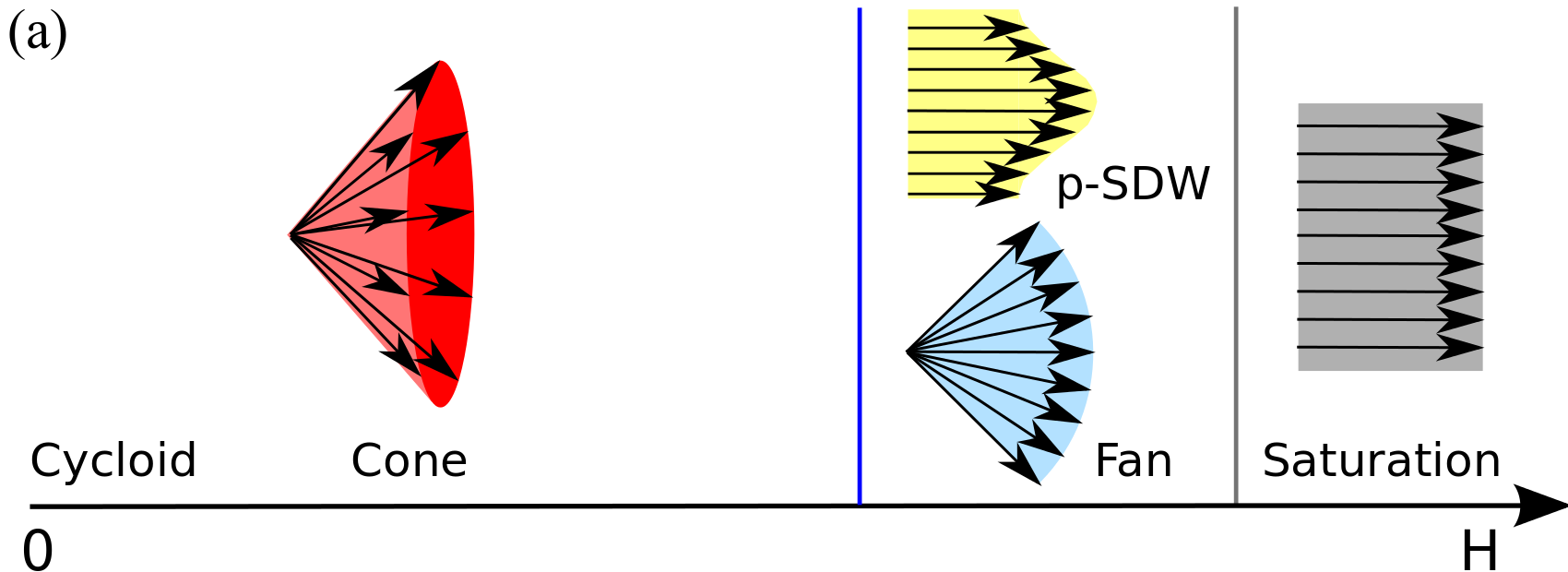}\\ \vspace{1mm}
\includegraphics[width=\columnwidth]{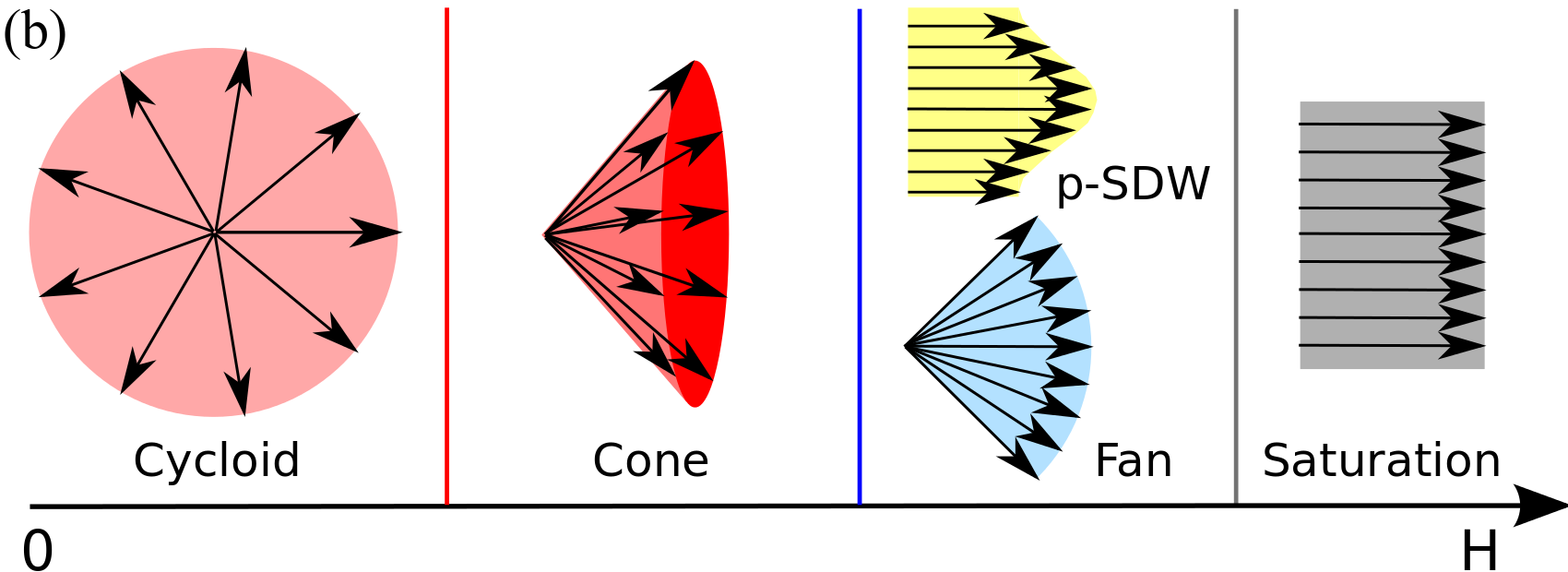}\\ \vspace{1mm}
\includegraphics[width=\columnwidth]{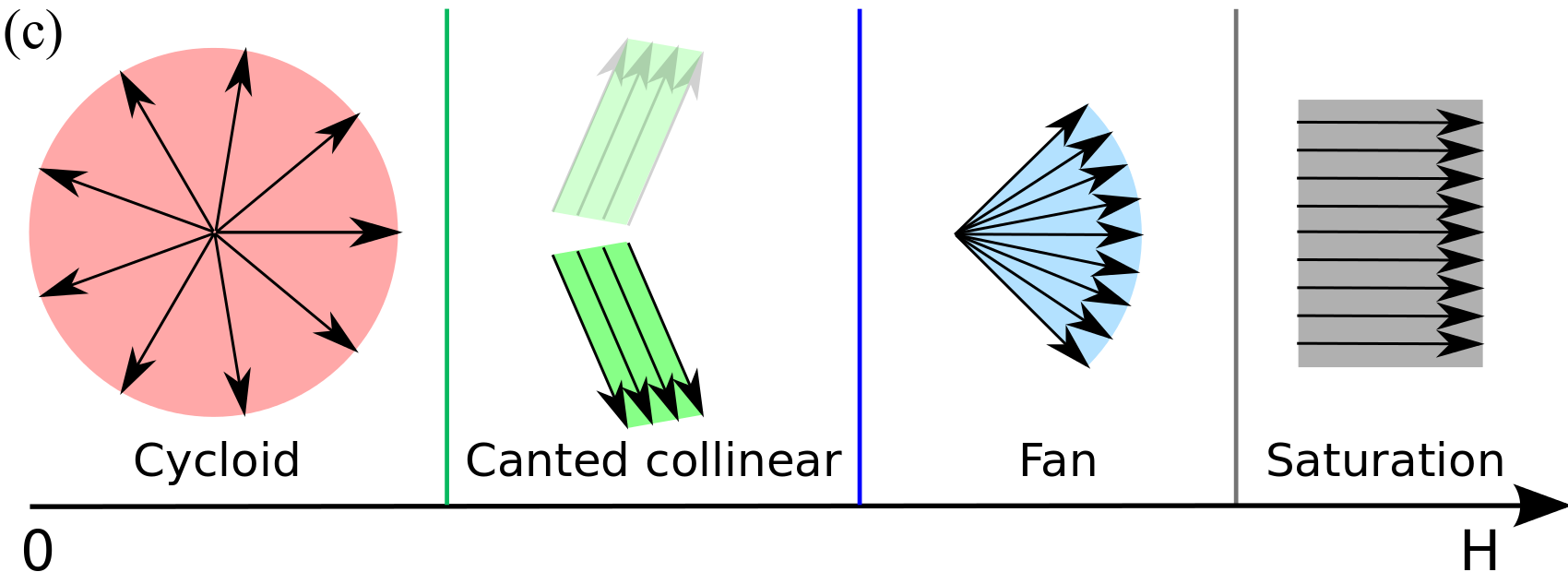}
\caption{Schematic phase sequence of quasi-one dimensional spin arrays with NN ferromagnetic and NNN antiferromagnetic intrachain exchange in the vicinity of the quantum critical point. In presence of interchain interactions, p-SDW stands for $p$-type quantum spin-density wave states or spin-multipolar bond order. 
SU(2)-symmetric exchange leads to the phase sequence (a) with a p-SDW phase for all magnetic field directions. U(1)-symmetric exchange of easy-plane type implies sequence (a) for a magnetic field along the symmetry axis (hard axis), and (b) for a field direction in the easy plane, with p-SDW-phases close to saturation ($p=2$). 
In linarite, we observe the phase sequence (a) with field along the hardest axis, (b) with field along the easiest axis, and (c) for field along the intermediate axis, with fan phases instead of p-SDW phases close to saturation. 
}
\label{phases}
\end{figure}

Linarite with chemical composition \lina\ crystallizes in the space group P2$_1$/m with $\beta=102.65\grad$ (see \cite{Bach61,Ara62,Eff87,Scho09} and Supplemental Material \cite{SM}). The crystal structure contains strongly buckled chains of edge-sharing CuO$_{4}$ plaquettes running along the monoclinic $b$ axis. Like many CuO$_{2}$-ribbon compounds, it orders at zero magnetic field into a cycloidal magnetic structure with incommensurate propagation vector {\bf k} and simultaneous ferroelectricity. In linarite the cycloid plane is roughly perpendicular to the CuO$_2$ chains and contains the $b$ axis \cite{Wil12}, while ${\bf k}=(0, 0.19, \frac{1}{2})$ and $T_N=2.8$~K \cite{Yas11,Pov16}. For convenience we introduce an orthogonal $xyz$-spin coordinate system, where $x$ and $y$ are in the spin-cycloid plane with $y||b$ and $z$ is perpendicular to the cycloid plane.
The magnetic field--temperature ($H$--$T$) phase diagram has been studied with various techniques for magnetic field directions parallel and perpendicular to $b||y$ \cite{Wil12,Pov16,Scha13}, but not along $x$. The presence of only one zero-field transition implies an easy-plane type anisotropy as the largest deviation from isotropic exchange, leading to schematic phase diagrams as shown in Figs.~\ref{phases}(a) and \ref{phases}(b) and as observed in LiCuVO$_4$ \cite{Mou12,Mou11}. However, linarite exhibits an extended commensurate phase for $H||b$ \cite{Wil12,Wil16} as in Fig.~\ref{phases}(c), in striking resemblance to the mean-field phase diagram of MnWO$_4$ \cite{Gvo16}. For field directions perpendicular to $b$, the commensurate phase appears to be absent \cite{Wol12,Scha13,Pov16}. This indicates the substantial role of anisotropy in linarite, in stark contrast to recent attempts to model the high-field behaviour \cite{Wil16} and zero-field spin dynamics \cite{Rul17} of linarite based on an entirely isotropic Hamiltonian.

\begin{figure}
\includegraphics[width=0.5\textwidth]{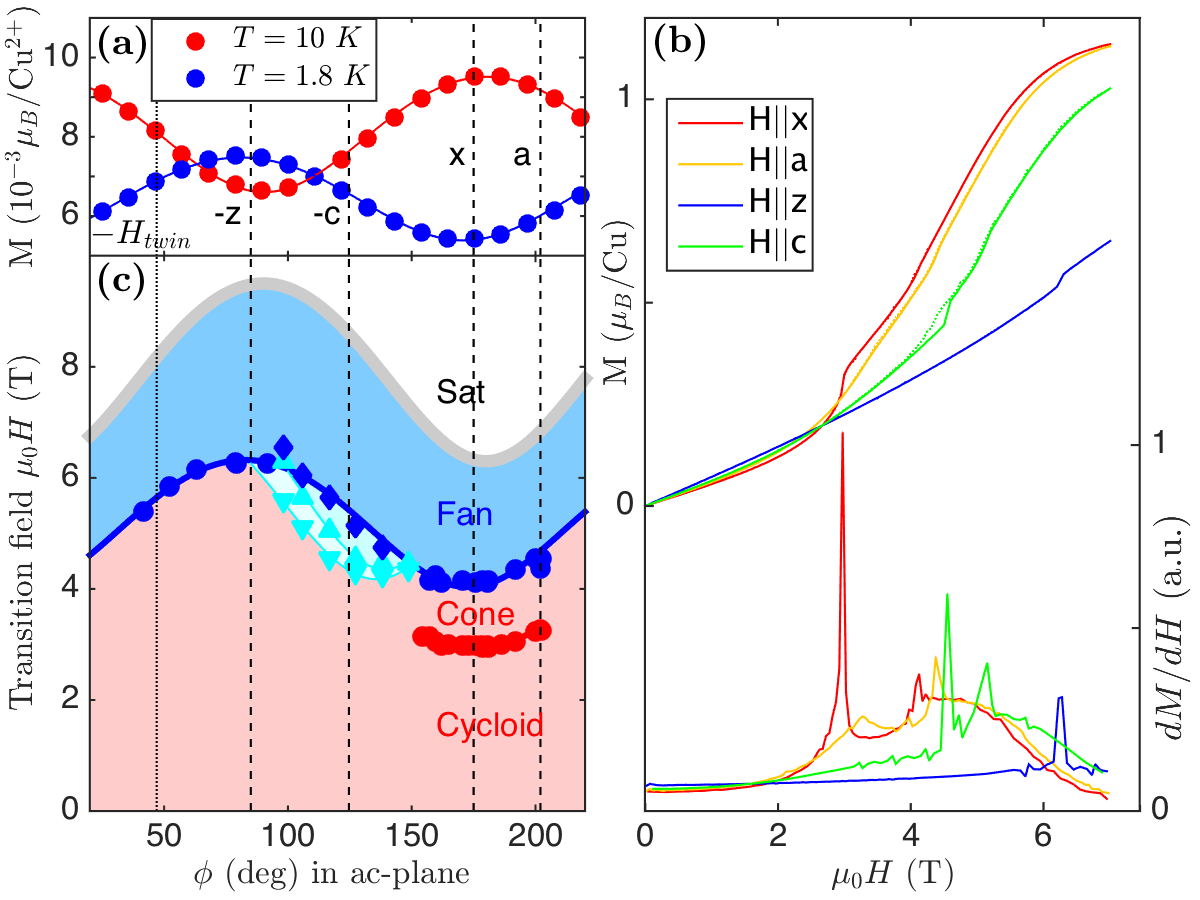}
\caption{(a) Magnetization $M$ as a function of rotation angle in the $ac$ plane at $T=9.995(1)$~K and $T=1.830(3)$~K for $\mu_0 H=100$~Oe. The longest axis of the $g$-tensor (maximum of $M$ at 10~K) and the spin-plane axis $x$ are less than 10$^{\circ}$ apart. 
(b) Representative  $M(H)$ curves at $T=1.8$~K, with their derivatives. Solid curves indicate raising field, dotted lowering field. 
(c) Phase diagram as a function of rotation angle in the $ac$ plane and field strength $\mu_0 H$ at $T=1.8$~K. The spin-flop transition exists only in a narrow angular region, with minimal field strength for the field direction parallel to the spin plane, $H||x$.}
\label{figmag}
\end{figure}

Magnetization as a function of the rotation angle around $b$ with the magnetic field direction in the $ac$ plane was measured on an untwinned crystal (see Ref.\ \cite{SM}) at $T=10$~K, i.e. above the ordering temperature, and at $T=1.8$~K in the cycloid phase, see Fig.~\ref{figmag}(a). 
The angular phase shift between $T=4$~K and 10~K in our data is negligible, $1\pm 1^\circ$, as it is in the paramagnetic phase for temperatures between 5 and 25~K \cite{Szy09}.
Hence, the $T=10$~K data reveal the approximate direction of the principal axes of the $g$ tensor perpendicular to $b$.  At $T=1.8$~K, in the cycloid phase, the susceptibility is dominated by the spin correlations  rather than the local $g$ tensor, and the magnetization is lowest for a field direction in the cycloid plane. Our data show that the cycloid axis $x$ is very close to the longest axis of the $g$ tensor, see Fig.~\ref{figmag}(a). The orientation of the spin-plane agrees well with the electric polarisation \cite{Pov16} and with unpolarised neutron diffraction \cite{Wil12}.

Magnetization measurements as a function of magnetic field for different directions in the $ac$ plane are shown in Fig.~\ref{figmag}(b) and the corresponding transitions are shown in Fig.~\ref{figmag}(c). For field directions approximately orthogonal to the cycloid plane, only one transition is observed below saturation: it corresponds to a kink in $M(H)$ without hysteresis, see $H||z$ in Fig.~\ref{figmag}(b).
In an intermediate angular range including the $c^{\ast}$ and $c$ directions, this transition splits into two, the lower with a large hysteresis, see $H||c$ in Figs.~\ref{figmag}(b) and \ref{figmag}(c). These findings agree well with the previously studied field directions in magnetization and dielectric measurements \cite{Scha13,Pov16}. As the field direction approaches the spin-plane, the small additional phase narrows and disappears, see the cyan region in Fig.~\ref{figmag}(c). With the field direction in the spin-plane, $H||x$, the upper phase transition has its minimum transition field of 4.1~T, and the saturation is reached above about 6.3~T, see $H||x$ in Fig.~\ref{figmag}(b). Close to the spin-plane, an S-shaped hysteresis-free magnetization step reveals a new low-field phase transition with a minimum transition field of 3~T when the field is parallel to the spin-plane $x$. This spin-flop transition (see below) has not been reported so far. It exists for field directions up to about $\pm$30\grad\ out of the spin plane, see Fig.~\ref{figmag}(c), including the crystallographic $a$ direction. At the angular end-points, the transition field reaches 3.3~T.

\begin{figure}
\includegraphics[width=0.9\columnwidth]{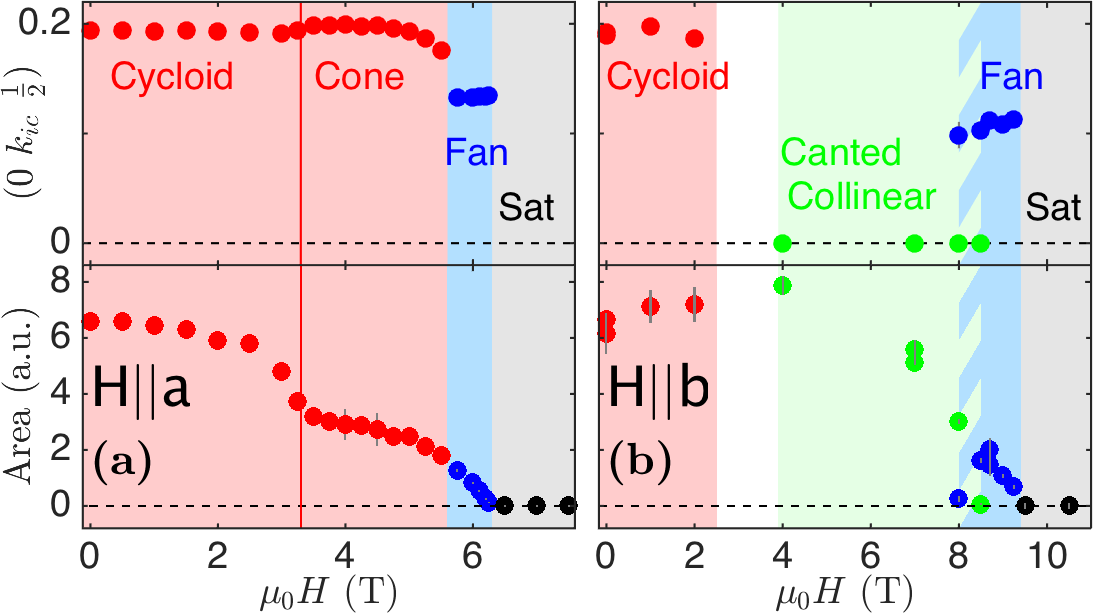}
\includegraphics[width=0.9\columnwidth]{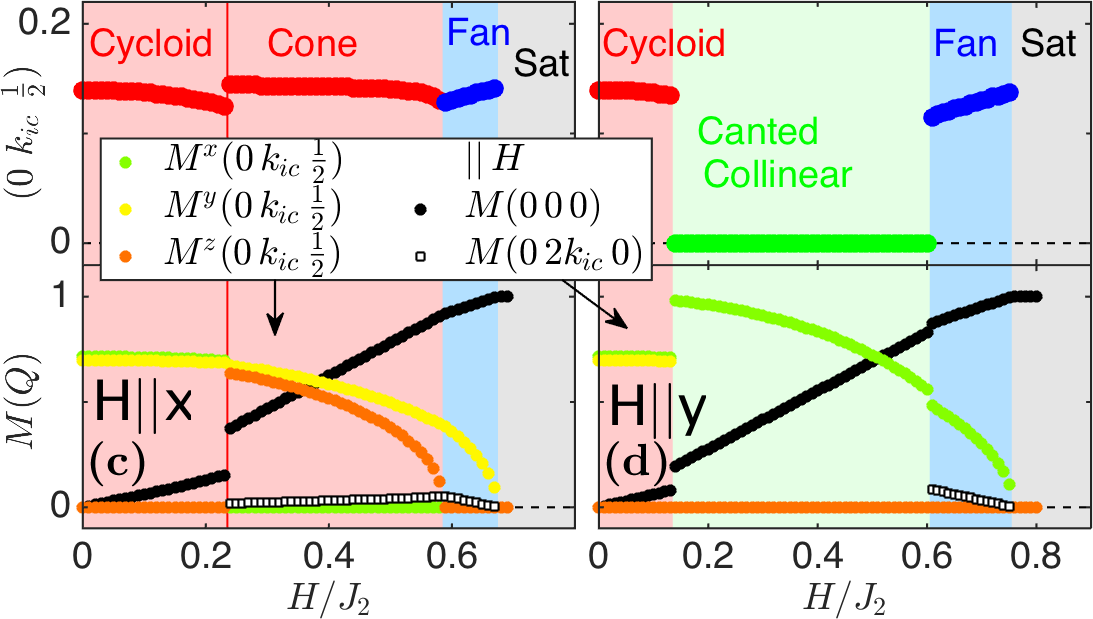}
\caption{(a--b) Measured ordering wave vector $(0,k_{\rm{ic}},\frac{1}{2})$ and integrated Bragg peak intensity at $T=60$~mK as a function of  magnetic field along (a) $H||a$ and  (b) $H||b$. The   integrated Bragg peak intensity at $\bm Q$ is proportional to $|\bm M(\bm Q)-\bm M(\bm Q)\hat{\bm Q}|^2$.  
(c--d) Calculated ordering wave vector $(0,k_{\rm{ic}},\frac{1}{2})$, uniform $\bm M(0,0,0)$ and staggered magnetization amplitudes $\bm M(\bm Q)$ at $T=0$ as a function of the reduced magnetic field $H/J_2$ along (c) $H||x$ (near the $a$ axis) and (d) $H||y$ ($||b$).
}
\label{figkic}
\end{figure}

Neutron diffraction \cite{SM} with $H||a$ at a temperature of 60~mK shows that the propagation vector remains incommensurate $(0,k_{\rm{ic}},\frac{1}{2})$ from zero field up to the saturation field, $H_{\rm sat}$, see Fig.~\ref{figkic}(a). At the $H\approx 3.3$~T transition, we observe the typical signature of a spin-flop: $k_{\rm{ic}}(H)$ jumps slightly upward, while the magnetic Bragg intensity decreases towards the transition, precisely as observed near the spin-flop transitions of LiCuVO$_4$ \cite{Mou11}. Neutron diffraction is sensitive to the staggered magnetic moment perpendicular to the wave vector, here $\bm Q=(0,k_{\rm{ic}},\frac{1}{2})$, which is close to $c^{\ast}$.  At the spin-flop transition, $H \approx 3.3$~T, the fully visible $a$-component of the cycloid flips to $c^{\ast}$ in the cone phase, where it is nearly invisible, while the staggered $b$ component remains essentially unchanged. This leads to the considerable drop in the Bragg peak intensity at the spin-flop transition.
The second phase transition at $H\approx 5.5$~T is preceded by a decrease of $k_{\rm{ic}}(H)$ with growing $H$, and followed by increasing $k_{\rm{ic}}(H)$ above the phase transition in contrast to the decreasing $k_{\rm{ic}}(H)$ expected for a spin-multipolar quantum phase. The intensity displays no visible anomaly. This means that the visible component of the staggered moment, $m_{b}$, remains practically unchanged across the transition. The staggered moment vanishes at about $H=6.3$~T, while the uniform moment (measured via the intensity of the (200) reflection, not shown) saturates, in agreement with the saturating magnetization, Fig.~\ref{figmag}(b). Complementary magnetization and neutron diffraction measurements at $T\sim 1.5$~K confirm that the spin-flop transition field is nearly temperature-independent, while the 5.5-T transition shifts to lower fields and at $T=1.8$~K corresponds to the blue line in Fig.~\ref{figmag}(c).

For $H||b$, our neutron diffraction data at $T=60$~mK confirm the major phase sequence observed in magnetization and neutron diffraction at higher temperatures  \cite{Wil12,Wil16}. This is illustrated in Fig.~\ref{figkic}(b): $k_{\rm{ic}}(H)$ jumps from incommensurate to commensurate values and back to incommensurate. In the phase just below saturation we observe -- precisely as for $H||a$ -- a continuous increase of $k_{\rm{ic}}(H)$ with $H$ at constant temperature of 60~mK, in sharp contrast to the decrease $k_{\rm{ic}}\propto (1-m)/p$ predicted for spin-multipolar quantum phases, and observed in LiCuVO$_4$ \cite{Mou12}.

\begin{figure}
\includegraphics[width=\columnwidth]{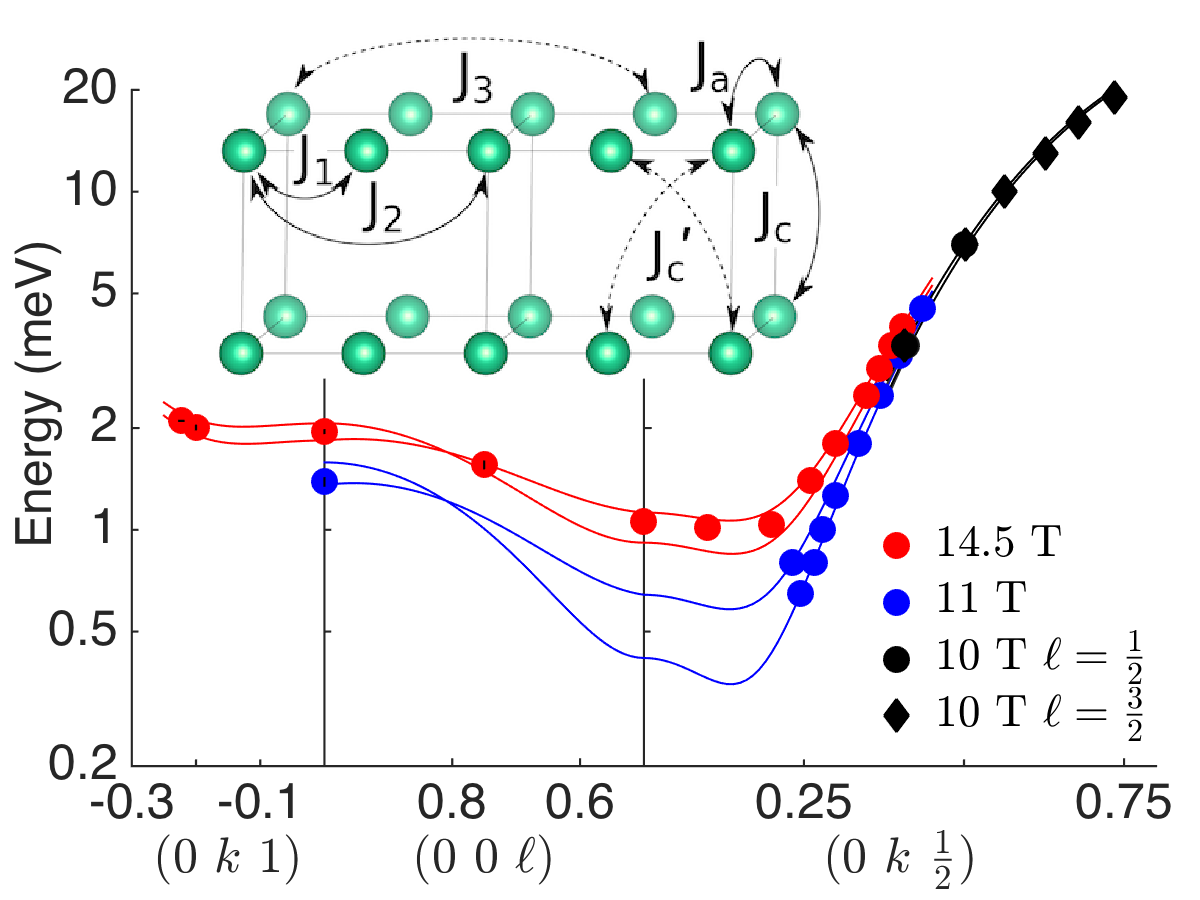}
\caption{Dispersion of linarite for $H||a$ with $H>H_{sat}$ (symbols) and spin-wave fit (lines) to the two twins. The inset displays the interaction scheme with Cu$^{2+}$ in two unit cells in the $b$-direction. Relevant interactions are shown as solid lines, additionally considered interactions as dashed lines.}
\label{figdis}
\end{figure}

The observation of entirely different phase sequences in three orthogonal field directions $x,y,z$ demonstrates that 
a bi-axial anisotropy is essential to understand the magnetism of linarite. We therefore introduce a minimal spin-$1/2$ Hamiltonian 
\begin{eqnarray}
{\cal H} &=& J_1 \sum_{\langle ij\rangle}^{\rm NN} \Bigl[ (1+\delta) S_{i}^{x} S_{j}^{x}  + S_{i}^{y} S_{j}^{y} + (1-\epsilon) S_{i}^{z} S_{j}^{z} \Bigr] + \nonumber\\
&+&J_2 \sum_{\langle ij\rangle}^{\rm NNN} {\bm S}_{i} {\bm S}_{j} + J_c \sum_{\langle ij\rangle}^{\mbox{\tiny inter}} {\bm S}_{i} {\bm S}_{j}  
- \mu_B \sum_{i,\alpha} g_{\alpha\alpha} H^{\alpha} S_i^{\alpha}\quad\quad \label{Ham}
\end{eqnarray}
with orthorhombic anisotropy included on the strongest ferromagnetic NN bonds $J_1<0$. 
A significant second-neighbor coupling $J_2>0$ beyond the critical ratio $J_2/J_1 = -1/4$ produces the incommensurate spiral structure along the chain $b$ direction, whereas  $J_c>0$ between direct neighbors in the $c$ direction, see Fig.~\ref{figdis}, assures an antiferromagnetic spin arrangement in this direction with an overall propagation  wave vector $(0,k_{\rm{ic}},\frac{1}{2})$, as observed in linarite.
We further neglect a mismatch between the principal axes of the $g$ tensor and the principal axes of the main exchange, since our magnetization measurements above and below $T_N$ show that the mismatch is small, see Fig.~\ref{figmag}(a). The antisymmetric Dzyaloshinski-Moriya interactions can play only a subsidiary role, see \cite{SM}. 

We investigate the competition between incommensurability and anisotropy described by (\ref{Ham}) using real-space mean-field simulations \cite{Gvo16}
at zero temperature, which for $S=1/2$ is equivalent to the classical minimization. 
This approach is justified by the large size of the ordered magnetic moments in linarite in zero magnetic field, 1.05$\mu_B$ \cite{Wil12}, which
indicates the suppression of quantum effects by anisotropy and interchain coupling. 
Further details of our approach are given in \cite{SM}. The general qualitative behavior of the equilibrium magnetic structure described by (\ref{Ham}) is as follows. In zero field, an easy-plane anisotropy $\epsilon>0$ stabilizes the cycloid in the $xy$ plane  
with a propagation vector $\cos(\pi k_{\rm{ic}}) = -J_1/(4J_2)$. The in-plane magnetic field ${\bf H}\parallel x,y$ induces a spin-flop transition into a distorted
conical state.  For small $k_{\rm{ic}}$, a small in-plane anisotropy $\delta>0$ is sufficient to suppress the spin-flop for ${\bf H}\parallel y$ replacing it with a transition into a canted commensurate state with the staggered moments oriented along the easy axis $m(0,k_{\rm{ic}},\frac{1}{2})\parallel x$.
For $H||z$, the cone phase is stable from zero field to near saturation.
For all three field orientations, the saturated phase is preceded by an incommensurate fan phase. The absence of rotational symmetry in the spin Hamiltonian
(\ref{Ham}) leads to variations of the propagation vector in a magnetic field, which are most pronounced in the fan state, corroborating previous theoretical work \cite{Nag67}.

Detailed comparison with the experimental results shown in Fig.~\ref{figkic} have been made for a set of parameters relevant for linarite: $|J_1^{yy}/J_2|=3.62$, $|J_c/J_2|=0.17$, $\epsilon=0.01$, $\delta=0.005$, and $g=2$. The obtained phase sequences match the major observed phases for all three field directions  sketched in Fig.~\ref{phases}. The calculation reproduces correctly numerous qualitative features, such as the variation of the wave vector as a function of the magnetic field or the type of anomaly of the staggered and uniform magnetizations at the phase transitions (see also \cite{SM}). 

The excellent qualitative agreement between theory and experiment, and notably the increasing $k_{\rm{ic}}(H)$ with field, identifies the high-field phases of linarite as fan phases, in contrast to spin-multipolar quantum phases that should feature a decreasing $k_{\rm{ic}}\propto (\frac{1}{2}-m)/p$. Since $k_{\rm{ic}}(H)$ increases with field close to saturation for two perpendicular field directions in the cycloid plane at a {\it constant} temperature of $T=60$~mK, we can rule out a succession of spin-multipolar phases \cite{Wil16}. The latter was suggested on the basis of measurements of wave vectors $k_{\rm{ic}}(H)$ along lines of varying and rather elevated temperatures (i.e., measurements of $k_{\rm{ic}}(H,T)$ rather than $k_{\rm{ic}}(H,T\!\rightarrow\!0)$, see figure 2b in \cite{Wil16} and figures 5.49 and 5.16 in \cite{Wil13}).

In order to establish how close linarite is to this quantum critical point,
we have determined the exchange interactions from INS measurements at low temperatures of the spin-wave spectrum above $H_{\rm sat}$, where the magnetic moments are aligned parallel to the magnetic field, and hence classical spin-wave theory is applicable. The twinning observed on all large natural crystals of linarite was taken into account explicitly, see \cite{SM}.

A global fit of all data taken at different field strengths (10~T, 11~T, 14.5~T) gives $J_1=-14.5(2)$~meV, $J_2=3.93(6)$~meV, direct interchain exchange $J_{c}=0.7(1)$~meV,  vanishing $J_a=-0.1(2)$~meV and diagonal interchain exchange $J_{c}^\prime=-0.06(3)$~meV, $g_a=2.5(3)$, $g_{twin}=2.3(3)$, $\epsilon=0.03(1)$, and $\delta < 0.005$. Our data rule out sizable diagonal interchain exchange ($J_{c}^\prime\sim -0.1 J_1$) that had been supposed in previous interpretations of $k_{\rm{ic}}(H,T)$ \cite{Wil16}, while the  low-energy zero-field dispersion shown in \cite{Rul17} is compatible with a wide range of parameters and cannot distinguish between diagonal ($J_{c}^\prime$) and direct ($J_{c}$) interchain exchange. Including third-nearest neighbour intrachain exchange $J_3$ improves the fit slightly \cite{SM}. With or without $J_3$ linarite is very close to the ferromagnetic phase boundary in the parameter space, about an order of magnitude closer than e.g. LiCuVO$_4$.

In conclusion, we find linarite very close to the critical line where ferromagnetism sets in.
Our findings demonstrate that at such close proximity to the critical point, tiny anisotropies and interchain interactions are ``highly relevant''  and sufficient to suppress higher-order spin-multipolar quantum phases. In real materials, the stability of quantum spin-multipolar phases in competition with dipolar long-range order may therefore be more limited than suspected so far.
Our angular-dependent magnetization and our neutron diffraction measurements  establish three entirely different phase sequences in three orthogonal field directions $x,y,z$ and imply that the Hamiltonian describing linarite has at most orthorhombic symmetry, in excellent agreement with real-space mean-field calculations for spin 1/2. A quantitative description of the phase diagram, incommensurability, or details of the uniform or staggered magnetization are left for future investigations.
Our work shows that in the vicinity of the quantum critical point the fan phase occurs in the frustrated quantum chain and is stabilized by a tiny orthorhombic anisotropy.  Further away from the critical point, the frustrated nature of the interactions leads to a close competition of different ground states including different types of dipolar long-range order. Orthorhombic anisotropy can induce ordinary ($p=1$) spin-density waves and/or fan phases both as a function of temperature and as a function of magnetic field  \cite{Gvo16}, and can play a role  for  the long-range ordered spin-density wave structures observed e.g. in $\beta$-TeVO$_4$ in a broad range of temperatures at zero magnetic field  \cite{Pre15,Pre16,Wei16}.

\end{document}